\documentstyle[12pt]{article}

\newcommand{\EQ}{\begin{equation}}
\newcommand{\EN}{\end{equation}}

\newcommand{\bear}{\begin{eqnarray}}
\newcommand{\ear}{\end{eqnarray}}
\newcommand{\bt} { \begin{tabular} }
\newcommand{\et}{ \end{tabular} }
\newcommand{\bc} { \begin{center} }
\newcommand{\ec}{ \end{center} }

\newcommand{\btb} { \begin{table} }
\newcommand{\etb}{ \end{table} }

\begin{document}

\topmargin 0pt
\oddsidemargin 5mm
\newcommand{\NP}[1]{Nucl.\ Phys.\ {\bf #1}}
\newcommand{\PL}[1]{Phys.\ Lett.\ {\bf #1}}
\newcommand{\NC}[1]{Nuovo Cimento {\bf #1}}
\newcommand{\CMP}[1]{Comm.\ Math.\ Phys.\ {\bf #1}}
\newcommand{\PR}[1]{Phys.\ Rev.\ {\bf #1}}
\newcommand{\PRL}[1]{Phys.\ Rev.\ Lett.\ {\bf #1}}
\newcommand{\MPL}[1]{Mod.\ Phys.\ Lett.\ {\bf #1}}
\newcommand{\JETP}[1]{Sov.\ Phys.\ JETP {\bf #1}}
\newcommand{\TMP}[1]{Teor.\ Mat.\ Fiz.\ {\bf #1}}
     
\renewcommand{\thefootnote}{\fnsymbol{footnote}}
     
\newpage
\setcounter{page}{0}
\begin{titlepage}     
\begin{flushright}
IFTA-97-36
\end{flushright}
\vspace{0.5cm}
\begin{center}
\large{ On the integrability of the $SU(N)$ Hubbard model }\\
\vspace{1cm}
\vspace{1cm}
{\large $M.J. Martins$  } \\
\vspace{1cm}
{Instituut voor Theoretische Fysica, Universiteit van Amsterdam\\
Valcknierstraat 65, 1018 XE Amsterdam, The Netherlands}\\
\vspace{0.3cm}
{and}\\
\vspace{.3cm}
{Universidade Federal de S\~ao Carlos\\
Departamento de F\'isica \\
C.P. 676, 13560~~S\~ao Carlos, Brazil}\\
\end{center}
\vspace{0.5cm}
     
\begin{abstract}
We exhibit explicitly the intertwiner operator for the monodromy
matrices of the recent proposed $SU(N)$ Hubbard model \cite{MAS}. This produces
a new family of non-additive $R$-matrices and generalizes an earlier
result by Shastry \cite{SA}.
\end{abstract}
\vspace{.15cm}
\vspace{.1cm}
\vspace{.15cm}
\end{titlepage}

\renewcommand{\thefootnote}{\arabic{footnote}}

For many years the one-dimensional Hubbard model has generated a great 
deal of attention, both due to the integrability properties \cite{LW,SA}
and the promising role as a paradigm for
strongly correlated electrons  systems \cite{PW,ESS}. The one-dimensional
Hubbard model has 
always appeared as an unusual and  solitary example, outside from an integrable
family of exactly solvable models. Very recently, however, 
Maassarani \cite{MAS} 
proposed an integrable $SU(N)$ generalization of the Hubbard model by using
as building blocks the Boltzmann weights of a ``free-fermion'' $SU(N)$ 
spin chain \cite{MAP}. This author also constructed the Lax operator
of the two-dimensional ``covering'' vertex model whose transfer matrix
commutes with the $SU(N)$ Hubbard Hamiltonian. The purpose of this note
is to complement Maassarani's work, by presenting the quantum $R$-matrix
which intertwines two monodromy matrices of the $SU(N)$ ``covering''
Hubbard model. This not only guarantees that transfer matrices with different
spectral parameters commute but also paves the way for an algebraic Bethe 
Ansatz solution (see e.g. \cite{KO}). In 
other words, we are interested in finding the
$R$-matrix which solves the Yang-Baxter algebra
\EQ
\check{R}_{12}(\lambda,\mu) {\cal L}_{01}(\lambda) {\cal L}_{02}(\mu) =
{\cal L}_{02}(\mu) {\cal L}_{01}(\lambda) \check{R}_{12}(\lambda,\mu)
\EN
where 
${\cal L}_{0i}(\lambda) $ is the $SU(N)$ Lax operator proposed in ref. \cite{MAS}.

For this purpose, we follow closely the strategy devised by Shastry \cite{SA}
who solved similar problem in the special case of $N=2$. First we apply
a ``gauge'' transformation \cite{SA} on the original $SU(N)$ Lax operator
proposed by Maassarani \cite{MAS}, rewriting it as
\EQ
{\cal L}_{0i}(\lambda) = 
{\cal L}^{(1)}_{0i}(\lambda) {\cal L}^{(2)}_{0i} (\lambda) 
e^{ h(\lambda) \hat{\rho}^{(1)}_{0} \hat{\rho}^{(2)}_{0} \otimes I_i}
\EN

In this equation
${\cal L}^{(k)}_{0i}(\lambda)$, $k=1,2$ are two commuting
copies of the vertex operator of the  ``free-fermion'' $SU(N)$ model \cite{MAP}.
The interaction is via the ``azimuthal'' operator \cite{MAS}
$\hat{\rho}^{(k)} = 
\displaystyle \sum_{ l < N} ( e^{(k)}_{ll} -e^{(k)}_{NN} )$ while 
constraint $h(\lambda)$ satisfies the relation \cite{SA,MAS}
\EQ
\frac{\sinh[2h(\lambda)]}{\sin(2 \lambda)} =U  
\EN
where the constant $U$ plays the role of an on-site Coulomb interaction.

Here we would like to solve relation (1) and (2) for $N \geq 3$. As an
example 
let us start with the case $N=3$. For this model it is convenient to write
the $SU(3)$ ``free-fermion'' vertex operator as
\bear
 {\cal{L}}^{(k)}_{0i}(\lambda) = 
\pmatrix{
 a(\lambda) &0 &0 &0 &0  &0  &0  &0  &0      \cr
 0 &b(\lambda) &0 &1  &0  &0  &0  &0  &0      \cr
 0 &0 &0 &0 &0  &0  &a(\lambda)  &0   &0      \cr
 0 &1 &0 &b(\lambda) &0  &0  &0  &0  &0      \cr
 0 &0 &0 &0 &a(\lambda) &0  &0 &0 &0 \cr
 0 &0 &0 &0 &0  &b(\lambda)  &0  &1  &0  \cr
 0 &0 &a(\lambda) &0 &0  &0  &0  &0  &0  \cr
 0 &0 &0 &0 &0   &1  &0  &b(\lambda)  &0    \cr
 0 &0 &0 &0  &0  &0  &0  &0  &a(\lambda)   \cr }
\ear
where $a(\lambda)=\cos(\lambda)$ and $b(\lambda) =\sin(\lambda)$. For this
representation the interaction term reads
\bear
\hat{\rho}^{(k)} = 
\pmatrix{ 1 &0 &0 \cr
          0 &-1 &0 \cr
          0 &0 &1 \cr}
\ear

We recall that we have performed an appropriate canonical transformation
on the original expressions for 
$ {\cal{L}}^{(k)}_{0i}(\lambda) $ and
$\hat{\rho}^{(k)} $ of ref. \cite{MAS}. In this way the ``free-fermion''
operator becomes more symmetrical, helping us to better
sort out the many different entries of equation (1).
The next step is to make
an Ansatz for $\check{R}_{12}(\lambda,\mu)$, guided by the results found
by Shastry for $N=2$ \cite{SA}. This suggests us to seek an expression for
the $R$-matrix having the following form 
\EQ
\check{R}_{12}(\lambda,\mu)=
{\cal L}^{(1)}_{12}(\lambda-\mu)
{\cal L}^{(2)}_{12}(\lambda-\mu)
+f(\lambda,\mu)
{\cal L}^{(1)}_{12}(\lambda+\mu)
\hat{\rho}^{(1)}_1 
{\cal L}^{(2)}_{12}(\lambda+\mu)
\hat{\rho}^{(2)}_1 
\EN

The coupling $f(\lambda,\mu)$ is determined via a brute force analysis, i.e.
by inserting the above Ansatz and the expression for the Lax
operator in the the Yang-Baxter algebra (1). Here one can use useful 
identities between the ``free-fermion'' Lax operator, and for instance
we have
\EQ
 {\cal{L}}^{(k)}_{12}(\lambda-\mu)
 {\cal{L}}^{(k)}_{13}(\lambda)
\hat{\rho}^{(k)}_{1} 
 {\cal{L}}^{(k)}_{23}(\mu)
\hat{\rho}^{(k)}_{2}  =
 {\cal{L}}^{(k)}_{23}(\mu)
\hat{\rho}^{(k)}_{2} 
 {\cal{L}}^{(k)}_{13}(\lambda)
\hat{\rho}^{(k)}_{1} 
 {\cal{L}}^{(k)}_{12}(\lambda-\mu)
\EN
\bear
 {\cal{L}}^{(k)}_{12}(\lambda-\mu)
 {\cal{L}}^{(k)}_{13}(\lambda)
 {\cal{L}}^{(k)}_{23}(\mu)
\hat{\rho}^{(k)}_{2} & =&
\frac{\sin(\lambda-\mu)}{\cos(\lambda+\mu)} \tan(\mu)
 {\cal{L}}^{(k)}_{23}(\mu)
 {\cal{L}}^{(k)}_{13}(\lambda)
 {\cal{L}}^{(k)}_{12}(\lambda+\mu)
\hat{\rho}^{(k)}_{1} 
\nonumber\\
&& \frac{\tan(\mu)}{\tan(\lambda)}
 {\cal{L}}^{(k)}_{23}(\mu)
 {\cal{L}}^{(k)}_{13}(\lambda)
\hat{\rho}^{(k)}_{1} 
 {\cal{L}}^{(k)}_{12}(\lambda-\mu)
\nonumber\\
&&
\frac{\sin(\lambda-\mu)}{\cos(\lambda+\mu)}\cot(\lambda)
 {\cal{L}}^{(k)}_{23}(\mu)
\hat{\rho}^{(k)}_{2} 
 {\cal{L}}^{(k)}_{13}(\lambda)
\hat{\rho}^{(k)}_{1} 
 {\cal{L}}^{(k)}_{12}(\lambda+\mu)
\hat{\rho}^{(k)}_{1} 
\ear
\bear
 {\cal{L}}^{(k)}_{23}(\mu)
\hat{\rho}^{(k)}_{2} 
 {\cal{L}}^{(k)}_{13}(\lambda)
\hat{\rho}^{(k)}_{1} 
 {\cal{L}}^{(k)}_{12}(\lambda+\mu)
\hat{\rho}^{(k)}_{1} &=&
\frac{\cos(\lambda+\mu)}{\cos(\lambda-\mu)} \left [
 {\cal{L}}^{(k)}_{23}(\mu)
\hat{\rho}^{(k)}_{2} 
 {\cal{L}}^{(k)}_{13}(\lambda)
 {\cal{L}}^{(k)}_{12}(\lambda-\mu)
\right.
\nonumber\\
&& \left. 
 {\cal{L}}^{(k)}_{23}(\mu)
 {\cal{L}}^{(k)}_{13}(\lambda)
\hat{\rho}^{(k)}_{1} 
 {\cal{L}}^{(k)}_{12}(\lambda-\mu) \right ]
\nonumber\\
&&- {\cal{L}}^{(k)}_{23}(\mu)
 {\cal{L}}^{(k)}_{13}(\lambda)
 {\cal{L}}^{(k)}_{12}(\lambda+\mu)
\hat{\rho}^{(k)}_{1} 
\ear

Now we can put all these information together, by expanding the exponential part
of the 
Lax operator (2), and by considering the above identities and its variants.
After some algebraic manipulations  
we find that a necessary condition for this function is
\EQ
f(\lambda,\mu) \sin(\lambda +\mu) \left [ \tanh[h(\lambda)] \tanh[h(\mu)] +1 
\right ] = \sin(\lambda -\mu) \left [
 \tanh[h(\lambda)] + \tanh[h(\mu)] \right ]
\EN

Keeping in mind the constraint (3), it is not difficult to find that the
above equation is satisfied by the following expression
\EQ
f(\lambda,\mu) = \frac{\cos(\lambda -\mu)}{\cos(\lambda+\mu)} \tanh[ h(\lambda)-h(\mu) ]
\EN

Remarkably enough, the solution for function $f(\lambda,\mu)$ is precisely 
the one found by Shastry for $N=2$. This means that the functional 
expression for the $R$-matrix is indeed universal, and the structure
of the Boltzmann weights depends only on a given representation for the 
``free-fermion'' 
building blocks. To verify this fact we have performed extensive checks 
for $N=3,4$ and we found that solution 
(11) is also a sufficient condition. This
leads us to conjecture 
that this should be valid for arbitrary integer $N \geq 2$.
All this resembles much the Baxterization procedure, and it seems that
a better understanding of these results
should necessarily start with the discovery of the
algebraic structure, such as Temperley-Lieb and braid-monoid, which is
behind the $R$-matrix construction (6) and (11). This could also prompt 
new representations for the building blocks, leading us to generate many
more integrable
``Hubbard models ''.

In summary we have exhibit the $R$-matrix for the recent proposed $SU(N)$
Hubbard model. We also verified, for $N=3,4$, that the $R$-matrix solution
satisfies the Yang-Baxter equation, by following the approach of Shiroishi and
Wadati \cite{SHWA}. We find that the amplitudes for the ``tetrahedral''
Zamolodchikov algebra is precisely 
the same of that found for 
$N=2$ \cite{SHWA},corroborating the universal 
scenario proposed above. We hope that such
$R$-matrix solution can be used together with the framework developed in ref.
\cite{PM} to solve the $SU(N)$ Hubbard model via an algebraic Bethe Ansatz
method.

\end{document}